\documentclass[12pt]{article}
\usepackage{amsfonts,amsmath,amssymb,amsthm,bbold}
\usepackage{scalerel}[2016/12/29]
\usepackage{epsfig,graphicx}
\usepackage{tikz}
\usepackage{float}
\usepackage{braket}
\usepackage{url,hyperref}
\usetikzlibrary{decorations,arrows}
\usetikzlibrary{decorations.markings}
\usetikzlibrary{shapes.geometric}

\newcommand{\bb}{\hskip -0.1cm}

\def\bb{\hskip -0.5mm}
\def\be{\begin{equation}}
\def\ee{\end{equation}}
\def\bea{\begin{eqnarray}}
\def\eea{\end{eqnarray}}

\def\half{{\textstyle {1\over 2}}}

\def\rme{\mathrm{e}}
\def\rmi{\mathrm{i}}

\def\rme{\mathrm{e}}
\def\rmi{\mathrm{i}}

\usepackage{pslatex}                 
\usepackage[frenchb]{babel} 
\NoAutoSpaceBeforeFDP
\usepackage{alltt}
\usepackage{color}

\usepackage{fancybox} 
\usepackage{amsmath,amsthm}

\definecolor{ctitre}{rgb}{.4,.5,.4}
\newcommand{\titre}[1]{
  \textcolor{ctitre}{\begin{center} \large \bf #1 \end{center}}
  \vspace{1ex minus 1ex}}

\definecolor{cro}{rgb}{1,0,0}                

\definecolor{cve}{rgb}{0,1,0}                

\definecolor{cbl}{rgb}{0,0,1}                

\definecolor{cbr}{rgb}{.8,.5,0}               

\def\rme{\mathrm{e}}
\def\rmi{\mathrm{i}}

\begin{document}

\titre{{{\Large \bf Inclusion statistics\footnote{Based on a talk given at the kick off meeting  "Intensive Period Quantum Mathematics @Polimi 2025"} 
}\\[0.5cm]}} 

\titre{{\bf St\'ephane Ouvry$^*$ and Alexios P. Polychronakos$^{**}$ } }

\titre{{\bf  $^*$  LPTMS ORSAY CNRS/Universit\'e Paris-Saclay}\\
{stephane.ouvry@universite-paris-saclay.fr}}
\vskip -1.2cm

\titre{{\bf  $^{**}$The Graduate Center and the City College of NY, CUNY}\\
apolychronakos@ccny.cuny.edu}
\vskip -1cm

\titre{{\Large   }}  

{\noindent{\bf Abstract:}
We present a historical review of anyon and exclusion statistics, introduced in the 1980s and 1990s respectively, and then turn to developments in
the recently introduced inclusion statistics. In contrast to exclusion statistics, where particles tend to be more exclusive than usual fermions, inclusion statistics particles tend
to be more gregarious than usual bosons and manifest an enhanced propensity to form condensates.
Inclusion and exclusion statistics are related through a duality transformation, generalizing the well-known Bose-Fermi duality. We conclude with a review
of the Calogero model realization of exclusion statistics and its extension to inclusion statistics.}

\noindent{\bf Introduction:}

 Let's start   from the  textbook Bose-Fermi statistical mechanics  for a general system with a 
 one-body spectrum made of $k$ discrete energy levels $\epsilon_1,\epsilon_2,\ldots,\epsilon_k$. As usual, we call $n_i$ the occupation number  of level $\epsilon_i$. The total number of particles  $N$, the $N$-body energy $E_N$  and the $N$-body partition function $Z_N$  read $N=\sum_{i=1}^{k}n_i, ~E_N=\sum_{i=1}^{k}n_i \epsilon_i$ and $Z_N=\sum_{E_N}\rme^{-\beta E_N}$. Standard manipulations reduce the  grand partition function $ Z=\sum_{N}z^NZ_N$ to 
\begin{align*}
 Z=\sum_{N}z^NZ_N =\sum_{N,E_N}z^N \rme^{-\beta E_N}
=& \sum_{n_1 \ldots n_k}z^{n_1+\ldots+n_k}(\rme^{-\beta\epsilon_1})^{n_1}\ldots(\rme^{-\beta\epsilon_k})^{n_k}\\
=&\sum_{n_1}(z\rme^{-\beta\epsilon_1})^{n_1}\ldots\sum_{n_k}(z\rme^{-\beta\epsilon_k})^{n_k}
\end{align*}
so that 
\begin{align*}
&  \text{Bose }\; n_i=0,1,\ldots,\infty &&\Rightarrow Z_0(z)=\big(\frac{1}{1-z\rme^{-\beta\epsilon_1}}\big)\ldots\big(\frac{1}{1-z\rme^{-\beta\epsilon_k}}\big)\\
& \text{Fermi } n_i=0,1 &&\Rightarrow Z_1(z)=\big(1+z\rme^{-\beta\epsilon_1}\big)\ldots\big(1+z\rme^{-\beta\epsilon_k}\big)
\end{align*}
Notice that
\be Z_0(z)=\frac{1}{Z_1(-z)}\label{1}\ee
where we introduced the notation $\{Z_{0}$,\;$Z_{1}\}$ for the Bose and Fermi grand partition functions.
Indeed we can write, e.g., for  the 2-body wavefunction,
\be
\psi(\vec r_1,\vec r_2)=\pm \psi(\vec r_2,\vec r_1) = \label{2}\exp(\rmi\pi g) \psi(\vec r_2,\vec r_1)\ee
where  $g$, the  statistical parameter,  takes the value  
 $0$ for Bose and $1$ for Fermi {statistics}. 
  
 We can interpret the well-known relation  (\ref{1}) as a sort of $g=1 \leftrightarrow 1-g=0$ duality, where the grand partition of bosons is obtained as the inverse of that
 of fermions provided that  $z$ is replaced by $-z$. {{We also note that this duality implies  for  the ensuing Bose and Fermi mean occupation 
 numbers\footnote{{From now on the additional subscript  for the mean occupation number $\langle n_i\rangle_{0,1}$  of level  $\epsilon_i$ 
 denotes the statistical parameter $g$, here $g=0$ or $g=1$.}} $\langle n_i\rangle_{0,1}$ of the energy level $\epsilon_i$
 \be \label{mean} \langle n_i\rangle_{{0,1}} ={z\rme^{-\beta\epsilon_i}{\partial \ln Z_{0,1}\over \partial z\rme^{-\beta\epsilon_i}}}={z\rme^{-\beta\epsilon_i}\over 1\mp z\rme^{-\beta\epsilon_i}}
 \ee
 that necessarily
 \be{\langle n_i\rangle}_{{0}}=-\langle n_i\rangle_{{1}}|_{z\to -z}\label{thisis}
 \ee}}

What is the  meaning, if any, of such a  duality ? Is it accidental   and  relevant only for Bose and Fermi statistics, or can it be  viewed  as  a particular case
of a more general scheme ? We will show that it  is indeed part of a more general relation between, on the one hand,   $g$-exclusion  statistics \cite{Haldane},
where the statistical parameter $g\ge 1$ is larger than 1, describing particles  more exclusive than ordinary fermions, and, on the other hand, a new kind of
quantum statistics called $(1\bb-\bb g)$-inclusion   \cite{bibi}, where the statistical parameter $1-g\le 0$ is negative, describing particles more gregarious than ordinary
bosons, and in particular  more easily condensable\footnote{Meaning at a higher critical temperature and a lower  critical dimension ($d=2$ rather than  $d=3$ as for ordinary bosons).} \cite{bibibis}.

\vspace{1cm}
\noindent{\bf To derive this general duality, a few standard and  not-so-standard  steps  are needed:}

First, it is  tempting to generalize (\ref{2}) by  considering  values of $g$ not confined to $0$ or $1$ but continuously interpolating  between them, i.e.,
$ 0\leq g\leq 1$.  It is, indeed, possible to do so, with the proviso that the  particles live in a 2d plane \cite{LM}. One then speaks of  anyon statistics
which does interpolate between Bose and Fermi statistics\footnote{More precisely,  in the anyon model  $g$ should be restricted by periodicity to the interval 
$[0,2]$, $g$=0 Bose, $g=1$ Fermi, $g=2$ Bose again.}.  The anyon statistical parameter $g$ can  be reinterpreted \cite{Wil} as the ratio 
$g=\phi/\phi_0$, where  $\phi$ is the flux of an infinitesimally thin flux tube carried by each particle, endowed also with an electric charge $e$, and
$\phi_0=h/e$ is the  associated flux quantum.  In this picture each  particle is coupled to the flux  of all  others via  quantum Aharonov-Bohm interactions, 
therefore  turning  ordinary, but interacting, bosons   into anyons with  non trivial statistics. Here the interactions are purely quantum with no classical counterpart, 
as it should for quantum statistics. Indeed  one can  equivalently see the particles  as free particles but with a multivalued wavefunction,
encoding their nontrivial braiding around each other, viz. the braid group, as illustrated in the figure below.

\begin{figure}[!htbp]
\vspace{1cm}
\hspace{-1cm}
\centering
\includegraphics[width=0.94\textwidth]{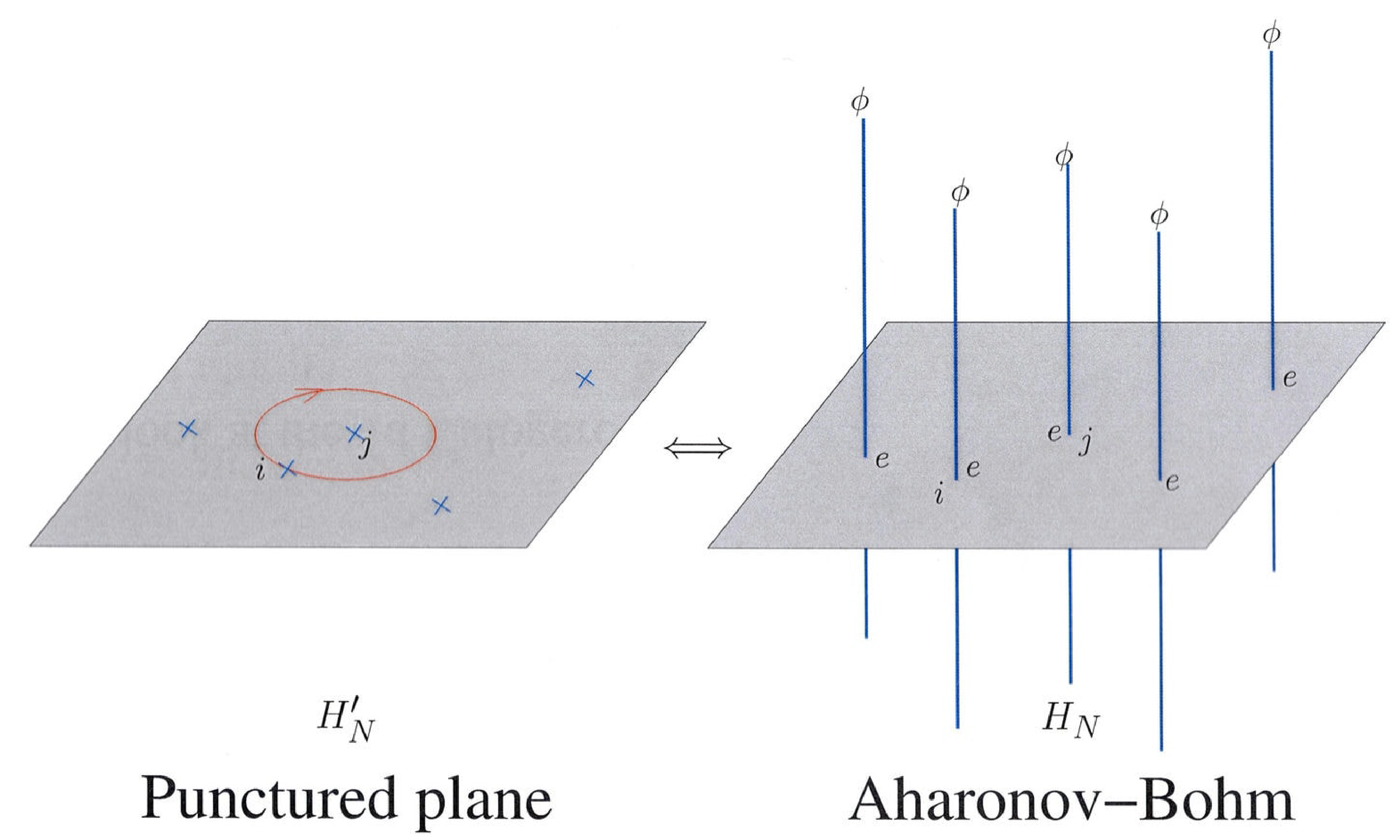}
\end{figure}

Second, it is known that the anyon model is notoriously resistant to unveiling its $N$-body energy spectrum. A  simplification is therefore needed.
To do so a standard and  natural step consists in adding a  constant magnetic field $B$ perpendicular to the plane, coupling the  charged particles
to it, and  projecting the system  onto the  lowest Landau level (LLL). One has moved from the anyon model to the  LLL-anyon model  \cite{Polyan,Trud,Das} 
which happens to be  solvable, meaning that a class of exact $N$-body eigenstates can be found interpolating between the   $g=0$ LLL-Bose basis  and the
$g=1$   LLL-Fermi basis\footnote{One reaches again by periodicity the Bose basis when $g=2$, but with a discontinuity due to $N$-anyon excited states joining  the LLL at this point (for a general review on the anyon and the LLL-anyon model see \cite{Ouvrybis}).}.
 From there,  the LLL-anyon thermodynamics \cite{Das}  can in turn be explicitly retrieved, with, e.g., a mean occupation number interpolating between the familiar Bose-Einstein and Fermi-Dirac mean occupation numbers.

Third,  one can go further by extending  LLL-anyon statistics  with  a statistical parameter $0\leq g\leq 1$   to statistics with a statistical parameter $g>1$.
To do so, one analytically continues the LLL-anyon  thermodynamic expressions to values
of the statistical parameter  $g$ above $1$. The end result is  called exclusion statistics, describing  particles more ``fermionic" than fermions,  i.e., 
more exclusive. It amounts to abandoning the original   picture of particles in the 2d plane endowed with a microscopic quantum anyon  Hamiltonian
and {considering the analytically continued   $N$-body LLL-anyon eigenstates and spectrum to higher $g$ for the purpose of counting states in the Hilbert space.}
This counting coincides with Haldane's exclusion statistics counting argument \cite{Haldane,Polex}.

 Finally, the last important and nontrivial step is to further analytically continue the statistical parameter  to {\it negative} values
 and call the resulting statistics ``inclusion statistics" \cite{bibi}.
This will lead to a general ``duality" transformation mapping $g$-exclusion statistics  with $g>1$ to
 $(1-g)$-inclusion statistics with $1-g<0$, the Bose-Fermi  duality $g=1\leftrightarrow 1-g=0$  in (\ref{1}) appearing as a special case. A  negative statistical parameter will also lead
 to nontrivial multiplicities in the $N$-body partition functions  encoding the effect of inclusion statistics beyond the Bose case.

Needless to say, we find it a bit surprising that nobody (to our knowledge) had considered the possibility of inclusion statistics ever since the
introduction of generalized statistics from the late 70's onwards.

Let us now describe the above steps in more detail. 

\vspace{0.2cm}
\noindent{\bf The  LLL-anyon  thermodynamics  in a nutshell: }  

One starts from the one-body LLL eigenstates 
\[z_i^{l_i}\rme^{-\frac{1}{2}\omega_c z_i \bar{z}_i}\]  where $ l_i=0,1,2,\ldots $ are  LLL angular momentum quantum numbers.
The LLL-spectrum at energy $\omega_c=eB/2$ has  
Landau degeneracy $k$ equal to  the flux of $B$ through the plane in units of $\phi_0$, which thus scales as  the (a priori infinite) area $S$ of the plane.
These one-body eigenstates allow in turn the construction of a class of exact  $N$-body LLL-anyon eigenstates,  here written in the singular gauge, i.e., for free particles with the multivalued wavefunction\footnote{The weak ordering of the $l_i$'s is the hallmark of Boson statistics; i.e., for $g=0$ the system is bosonic.
{For $g=1$, the Vandermonde prefactor makes the system fermionic.}}
\bea \nonumber \psi'(z_1, z_2, \ldots, z_N)&=& \;\prod_{i<j}{(z_i-z_j)}^{{g}}\;\;{{\rm Sym}}\;\;\prod_{i=1}^N {z_i}^{{l_i}}\rme^{-{1\over 2}{\omega_c}z_i\bar z_i}\quad\quad 0\le l_i\le l_{i+1} 
\eea
with the  degenerate  $N$-anyon LLL-spectrum $E_N = N\omega_c$. These $N$-body  LLL-anyon eigenstates   constitute a 
{complete} basis interpolating between the  LLL-Bose  basis and   the  LLL-Fermi basis when $g$ moves from $0$ to $1$, thereby solving the LLL-anyon model.

From the $N$-body LLL-anyon spectrum one can get the LLL-anyon thermodynamics \cite{Das}. 
The $N$-anyon partition functions $Z_N=\sum_{0\le l_1\le l_2\le ...\le l_N} \rme^{-\beta E_N}$
lead to the grand partition function $Z_g=\sum_{N=0}^{\infty} z^NZ_N$  from which  the cluster expansion $\ln Z_g=\sum_{n=1}^{\infty}  z^n\;{{c_g(n)}}$ follows. 
A one-body harmonic well potential $\half \omega^2 z\bar z$, introduced as a long distance regulator, lifts the infinite Landau degeneracy of the $N$-anyon spectrum
and endows it  with an explicit $g$-dependence. It  reads  
\bea
E_N &=& (\omega_t- \omega_c) (\sum_{i=1}^N l_i+g{N(N-1)\over 2}) +N \omega_t   \label{harmonic}\quad\quad{0\le l_i\le l_{i+1}}\eea where  
$\omega_t = \sqrt{\omega^2 + \omega_c^2}$. This nondegenerate spectrum in turn  allows 
 for the computation of otherwise  divergent or ambiguous expressions. 
An appropriate  $\beta\omega\to 0$ thermodynamic limit prescription, namely   ${1\over n(\beta\omega)^2}\to {S\over \lambda^2}$ where $\lambda$ is the thermal wavelength, then leads in the thermodynamic limit to
the LLL-anyon cluster coefficients {on} the plane
\be {c_g(n)= k{\rme^{-\beta n\omega_c}\over n}\prod_{j=1}^{n-1}{j-ng \over j}}\label{cluster}\ee
that {properly} scale as the LLL-degeneracy $k = B S /\phi_0$.
These coefficients are the key building blocks for the thermodynamics we are aiming to.
Indeed, {the cluster expansion leads to}
\[\ln Z_g=\sum_{n=1}^{\infty} z^n \; c_g(n)=k\ln y\]
where the effective single-level grand partition function $y$ satisfies
\be\label{eq}
y^g - y^{g-1} = z\rme^{-\beta\omega_c}
\ee
It follows that  the  LLL-anyon thermodynamics  is encoded in  
\be\label{thermo}\ln Z_g = k\ln y\;\;\;{\oplus}\;\;\;
y^g - y^{g-1} = z\rme^{-\beta\omega_c}\ee
 or equivalently
\be\label{bon}
\ln Z_g =\int_0^{\infty}\rho_{\text{LLL}}(\epsilon)\ln y\; d\epsilon\;\;\;{\oplus}\;\;\;
y^g - y^{g-1} = z\rme^{-\beta\epsilon}
\ee where 
$\rho_{\text{LLL}}(\epsilon)=k\;\delta(\epsilon-\omega_c)$ is the  LLL-density of states. 
The mean occupation number $ \langle n_{c} \rangle_g $ of a one-body LLL-level  then follows as 
\be
\nonumber \langle n_{c} \rangle_g =z{\partial \ln y \over \partial z} 
~~\Rightarrow~~ z\rme^{-\beta\omega_c}={\langle n_c\rangle_g\over \big(1+(1-g)\langle n_c\rangle_g\big)^{1-g}\big(1-g\langle n_c\rangle_g\big)^{g}}
\ee
reducing to (\ref{mean}) for $g=0,1$ (here the LLL-energy  $\omega_c$ replaces $\epsilon_i$).

\vspace{1cm} 
\noindent{\bf Exclusion statistics in a nutshell:}

From the above  one can go a step further: let $g$ now  become larger than $1$ and focus on the solution of (\ref{eq}) such that 
$y=1+z (\ldots)+\ldots$.   From (\ref{thermo}) one  obtains  
\be Z_g=y^{k}
=\sum_{N=0}^{\infty}z^N{\underbrace{\rme^{-\beta N\omega_c}{k(k+N(1-g)-1)!\over N!(k-Ng)!}}_{Z_N= N\text{-body LLL-anyon partition function} }}
\label{combina}\ee
The  $N$-body  LLL-anyon partition function  $Z_N$ appearing in the expansion above  tells us that when there are $N$ anyons in the LLL
their occupation degeneracy is given by the multiplicative factor ${k(k+N(1-g)-1)!/\big( N!(k-Ng)!\big)}$.
 An obvious question is: does this degeneracy admit a natural interpretation?
In fact, when $g$ is an integer, it counts the  number of ways to put $N$ particles in $k$ ordered levels such that each level is occupied by
 at most one particle and
there are at least $g-1$ unoccupied levels between any two occupied levels\footnote{Note that the counting obtained here is for levels arranged on a circle,
but this will be immaterial in the thermodynamic limit. An open lineal counting can be obtained at the price of minor modifications \cite{bibi}. Also,
the summation over $N$ in $Z_g$ extends to infinity, whereas it should obviously be truncated at $N=k/g$, which can be achieved by using
the other solutions of (\ref{eq}) \cite{bibi}.}.
This ``resistence" of particles to occupy nearby levels is the essence of $g$-exclusion statistics, which extends Pauli exclusion to a more exclusive rule for one-body level occupation \cite{Haldane},
as illustrated for $g=2$ with at least one empty level between any two occupied levels.

\begin{figure}
\hspace{1cm}
\includegraphics[width=0.95\textwidth]{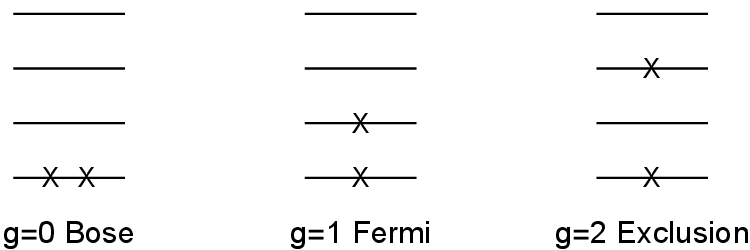}
\end{figure}

Quite generally, exclusion statistics arises by adopting this counting for a general one-body spectrum,
forgetting the LLL-anyon model from which it originates. This leads to a microscopic definition of exclusion statistics with levels
now corresponding to one-body eigenstates ordered by their energy.
In the thermodynamic limit, we can extend the concept of exclusion to fractional values of $g$. The multiplicative factor in (\ref{combina}) does not
admit any more a simple combinatorial interpretation, but we can accept it as an effective multiplicity for a number of states $k\gg1$. Partitioning the
energy spectrum to intervals $\Delta \epsilon$, each containing a large number of states, and applying (\ref{combina}) to each interval, leads to
a thermodynamical grand partition function given by (\ref{bon}) but now with $g$ fractional and $y$ representing an effective single-level grand
partition function. That is,
\be \ln Z_g =\int_0^{\infty}\rho(\epsilon)\ln y\; d\epsilon\;\;\;{\oplus}\;\;\;
y^g - y^{g-1} = z\rme^{-\beta\epsilon}\label{newg}\ee
describes particles with arbitrary $g$-exclusion statistics ($g>0$) and an arbitrary density of states $\rho(\epsilon)$.  The mean occupation number $\langle n_\epsilon \rangle_g$ of the energy level $\epsilon$ follows as
 \be
\label{meanmean} \langle n_\epsilon \rangle_g =z{\partial \ln y \over \partial z } 
~~\Rightarrow~~ z\rme^{-\beta\epsilon}={\langle n_\epsilon\rangle_g\over \big(1+(1-g)\langle n_\epsilon\rangle_g\big)^{1-g}\big(1-g\langle n_\epsilon\rangle_g\big)^{g}}
\ee
The cluster coefficients (\ref{cluster}), stripped of their Landau degeneracy factor $k$ and with $\omega_c$ replaced by $\epsilon$
\be {c_g(n)= {\rme^{-\beta n\epsilon}\over n}\,\prod_{j=1}^{n-1}{j-ng \over j}}\label{clusterbis}\ee
become the
effective single-level cluster coefficients for particles with $g$-exclusion statistics in a quantum state at energy $\epsilon$.

\vspace{0.5cm}
\noindent{\bf Moving on to inclusion statistics: }

Let us focus  on  these cluster coefficients in (\ref{clusterbis}): trading $g \to 1-g$ we notice
\[c_{1-g}(n)=(-1)^{n-1}c_g(n)\]
The effective single-level grand partition function $y$ in the cluster expansion $\ln y=\sum_{n=1}^{\infty}  z^n c_g (n)$
depends on $g$ and $z$, in addition to $\epsilon$, so from now on let's denote it $y_g(z)$. 
Trading $g \to 1-g$  amounts to
\[ \ln y_{1-g}(z)=
\sum_{n=1}^{\infty} c_{1-g}(n)z^n=
\sum_{n=1}^{\infty}(-1)^{n-1}c_{g}(n)z^n=
- \ln y_g(-z)\]
and thus\footnote{This can also be seen from $y^g - y^{g-1} = z\rme^{-\beta\epsilon}$ which can be rewritten as $(1/y)^{1-g}-(1/y)^{1-g-1}   = -z\rme^{-\beta\epsilon}$.}
\[y_{1-g}(z)=\frac{1}{{y_g(-z)}}\]
Upon using (\ref{newg}), this duality generalizes \cite{bibi} to the full partition function $Z_g (z)$
\be \label{yes} Z_{1-g}(z)=  \frac{1}{Z_g(-z)}\ee
so that
\be
\ln Z_{1-g} =-\int_0^{\infty}\rho(\epsilon)\ln y_g(-z)\; d\epsilon\;\;\;{\oplus}\;\;\;
y^g - y^{g-1} = z\rme^{-\beta\epsilon}\nonumber
\ee
It relates $g$-exclusion statistics with $g>1$ to $(1-g)$-inclusion statistics with statistical parameter  $1-g<0$ by inverting the $g$-exclusion grand partition
function and trading
$z\to -z$. We call the resulting statistics with a negative statistical parameter {\it inclusion} statistics, a term justified by the fact that, as we will see,
it corresponds to an {\it enhanced} multiplicity when more than one particles occupy $g$ adjacent levels, making them more gregarious than bosons. Eq.(\ref{yes}), then, is a duality relation between $g$-exclusion and ($1-g$)-inclusion.
This, in turn, implies for the mean occupation number $ {\langle n_{\epsilon}\rangle}_{1-g}$ of a level of  energy $\epsilon$
\be {\langle n_{\epsilon}\rangle}_{1-g} ={z{\partial \ln y_{1-g}\over \partial z}} ~~\Rightarrow~~
 z\rme^{-\beta\epsilon}={\langle n_{\epsilon}\rangle_{1-g}\over \big(1-(1-g)\langle n_{\epsilon}\rangle_{1-g}\big)^{1-g}\big(1+g\langle n_{\epsilon}\rangle_{1-g}\big)^{g}}
\label{meanmeanmean}\ee 
 that
 \be\langle n_\epsilon\rangle_{{1-g}}=-\langle n_\epsilon\rangle_{{g}}|_{z\to -z}\label{theend}\ee 
Clearly (\ref{yes},\ref{theend}) extend the Bose $\leftrightarrow$ Fermi duality (\ref{1},\ref{thisis}) to a general $g\leftrightarrow 1-g$ duality.

\vspace{0.5cm}
\noindent{\bf Inclusion statistics for a general discrete one-body spectrum $\epsilon_1,\epsilon_2,\ldots,\epsilon_k$:}

The exclusion-inclusion duality (\ref{yes})  has been derived in general for a system  in the thermodynamic limit with a  one-body density of states $\rho(\epsilon)$.
One can now declare it to apply to a general discrete spectrum $\epsilon_1,\epsilon_2,\ldots,
\epsilon_k$ as the one considered in the introduction. This will lead to the definition of {\it microscopic} (rather than thermodynamic) inclusion statistics.

To derive the properties of this microscopic inclusion statistics, start \cite{oulabis} from $g$-exclusion statistics for particles in this 
discrete one-body spectrum  with an integer statistical parameter $g>1$. By definition such particles are prevented to occupy levels
less than $g$ levels apart.  
The $n$-body partition function for such particles (warning: from now on the number of particles is denoted as $n$) is
\be\label{sala} Z_{n} = \sum_{i_1=1}^{k-gn+g} \sum_{i_2=1}^{i_1}\cdots \sum_{i_n=1}^{i_{n-1}}\rme^{-\beta \epsilon_{i_1+gn-g}}\ldots  \rme^{-\beta \epsilon_{i_{n-1}+g}}\rme^{-\beta \epsilon_{i_n}}\ee
It follows that the corresponding $g$-exclusion grand partition function for a gas of particles in the one-body spectrum $\epsilon_1,\epsilon_2,\ldots,\epsilon_k$ is
\[Z_{g} (z)=\sum_{n=0}^{(k+g-1)/g} z^{n} Z_n \]
with $Z_n$ given in (\ref{sala}).

According to our definition that the duality relation (\ref{yes}) also applies to a discrete one-body spectrum,
$(1-g)$-inclusion amounts to taking the inverse of this grand partition function with $z\to -z$, so that
\[\ Z_{1-g}(z)=\frac{1}{Z_g~(-z)}=\frac{1}{\sum_{n=0}^{(k+g-1)/g} (-z)^nZ_n}\]
is the $(1-g)$-inclusion grand partition function for a gas of particles in the same one-body spectrum $\epsilon_1,\epsilon_2,\ldots,\epsilon_k$.

Let's consider a few examples to see more precisely what happens:

\noindent $g=1\to 1-g=0$  (Bose-Fermi):
 \hspace*{-1.5cm}
 \[Z_0(z)\!=\!\frac{1}{Z_1(-z)}\!=\!\sum_{n_1 \ldots n_k=0}^{\infty}(z\rme^{-\beta\epsilon_1})^{n_1}\ldots(z\rme^{-\beta\epsilon_{k-1}})^{n_{k-1}}(z\rme^{-\beta\epsilon_k})^{n_k}\]
$g=2 \to 1-g=-1$:
 \hspace*{-1.5cm} 
 \begin{align*}  Z_{-1}(z)\!=\!\frac{1}{Z_2(-z)}\!=\!
\sum_{n_1 \ldots n_k=0}^{\infty}(z\rme^{-\beta\epsilon_1})^{n_1}{\binom{n_1+n_2}{n_1}}
\ldots
(z\rme^{-\beta\epsilon_{k-1}})^{n_{k-1}}{\binom{n_{k-1}+n_{k}}{n_{k-1}}}
(z\rme^{-\beta\epsilon_k})^{n_k}\nonumber\end{align*}
$g=3 \to 1-g=-2$ :
\hspace*{-1.5cm}
\begin{align*}
  Z_{-2}(z)\!=\!\frac{1}{Z_3(-z)}\!=\!
\sum_{n_1 \ldots n_k=0}^{\infty}&(z\rme^{-\beta\epsilon_1})^{n_1}{\binom{n_1+n_2+n_3}{n_1}}
\ldots
(z\rme^{-\beta\epsilon_{k-2}})^{n_{k-2}}{\binom{n_{k-2}+n_{k-1}+n_{k}}{n_{k-2}}} \\
&(z\rme^{-\beta \epsilon_{k-1}})^{n_{k-1}}{\binom{n_{k-1}+n_{k}}{n_{k-1}}}(z\rme^{-\beta \epsilon_k})^{n_k}
\end{align*}
The pattern here is that, while $g$-exclusion forbids particles to occupy levels less that $g$ apart, $(1-g)$-inclusion amounts to an enhanced multiplicity
when more than one particles occupy similarly neighboring levels. E.g., for  $g=2$, if $n_i=1$ then necessarily $n_{i\pm 1}=0$, whereas for
$1-g=-1$ one gets the enhanced occupation multiplicities ${\binom{n_i+n_{i\pm 1}}{n_i}}$.

\text{In general the $g$-exclusion  $\leftrightarrow (1-g)$-inclusion duality leads to } \be\label{itsnice}\nonumber Z_{1-g}(z)=\frac{1}{Z_g(-z)}=\sum_{n_1 \ldots n_k=0}^{\infty}{{{\text {m}}_{1-g}(n_1,\ldots,n_k)}}(z\rme^{-\beta\epsilon_1})^{n_1}\ldots(z\rme^{-\beta\epsilon_k})^{n_k}\ee
where the multiplicities ${\text {m}}_{1-g}(n_1,\ldots,n_k)\ge 1$  enhance the occupation weights of clusters of $g$ neighboring one-body quantum states, thereby  encoding $(1-g)$-inclusion. 
They are the product of multiplicities of clusters of $g$ adjacent states with the particles in each cluster considered
as distinguishable, divided by the corresponding multiplicities of overlaps between clusters. Defining 
\be
[i,j] := {{n_i + n_{i+1} + \cdots + n_j} \choose n_i \,,\, n_{i+1}\, ,\, \dots\, ,\, n_j} {~~\text{if}~~i<j, 
~~ [i,j] := 1 ~~\text{otherwise}}
\nonumber\ee
 for $k> g$ they read \cite{bibi} \be\label{mult}
{\text {m}}_{1-g} (n_1 ,\dots, n_k) =\; [1,g]\, [2,g+1] \cdots [k+1-g,k] \over [2,g]\, [3,g+1] \cdots [k+1-g,k-1]
\ee
{while for $k\le g$ they become the distinguishable particles multiplicities}
\be
{\text {m}}_{1-g} (n_1 ,\dots, n_k) =\; {{n_1 + n_2 + \cdots + n_k} \choose n_1 \,,\, n_2\, ,\, \dots\, ,\, n_k}
\nonumber\ee 
Finally, the  $(1-g)$-inclusion $n$-body partition functions directly follow as\footnote{Note that to avoid cumbersome notations the same symbol  $Z_n$ as in (\ref{sala}) for $g$-exclusion is used here for $(1-g)$-inclusion.}
\be\label{bibi} 
Z_{n}=\sum_{n_1, \ldots, n_k=0,\;\sum_{i=1}^k n_i=n
}^{n}{{\text {m}}_{1-g}(n_1,\ldots,n_k)}(\rme^{-\beta\epsilon_1})^{n_1}\ldots(\rme^{-\beta\epsilon_k})^{n_k}
\ee

The expressions (\ref{mult})  and   (\ref{bibi}) are surprisingly simple, given that we started from the $g$-exclusion $n$-body partition function (\ref{sala}),
then built the associated grand partition function, then inverted it and traded $z$ for $-z$, and we still end up with relatively simple and intuitive
expressions for the inclusion multiplicities and $n$-body  partition functions\footnote{
The mean occupation numbers  $\langle n_i\rangle_{g}$ of energy level $\epsilon_i$ for $g$-exclusion  i.e.,  the  rewriting of   (\ref{meanmean})  for a discrete spectrum  $\epsilon_1,\epsilon_2,\ldots,
\epsilon_k$,  are given by a recursion formula (for more details see \cite{oulabis}). Likewise the mean occupation numbers   ${\langle n_i\rangle}_{1-g}$  for ($1-g)$-inclusion i.e., the discrete rewriting of  (\ref{meanmeanmean}), are obtained via the duality relation  $\langle n_i\rangle_{{1-g}}=-\langle n_i\rangle_{{g}}|_{z\to -z}$ as specified in (\ref{theend}).}

\vspace{1cm}
\noindent{\bf The  Calogero model and its possible generalization: } 

Let's finally consider  the Calogero model  \cite{Calogero}, a paramount one-dimensional $n$-body quantum  model that realizes 
$g$-exclusion statistics \cite{Poly0,PolyP}.
In the presence of a one-body harmonic well potential $\frac{1}{2}\omega^2x^2$ as a long-distance regulator, the Hamiltonian is
\be\label{caloha}
H_n=-\frac{1}{2} \sum_{i=1}^n \frac{\partial^2}{\partial x_i^2} { - \sum_{i<j} \frac{g(1-g)}{\left(x_i-x_j\right)^2}}+\frac{1}{2} \omega^2 \sum_{i=1}^n x_i^2
\ee
where $g$ plays the role of the statistical parameter (for the integrability and eigenfunctions of the harmonic Calogero model see
\cite{PolyP,PolyX} ).
Its $n$-body  spectrum 
\be\label{top}
E_n=\omega \Big(\sum_{i=1}^n l_i+g \frac{n(n-1)}{2}+{n\over 2} \Big) ~,\quad\quad l_i \in  Z~,~~0  \leq l_i \leq l_{i+1}
\ee
is indeed a $g$-exclusion spectrum: defining the quasi-excitation parameters $l'_i=l_i+g(i-1)$, the $n$-body spectrum becomes
\be\label{stunning}
E_n=\omega\sum_{i=1}^n \Big(l_i^{\prime}+{1\over 2}\Big) ~, \quad\quad (i-1) g \leq l_i^{\prime} \leq l_{i+1}^{\prime}-g
\ee
The above looks like the spectrum of noninteracting particles with excitations $l_i'$. Note that the $l_i'$ differ by at least $g$,
but, for general $g$, are not integers. For integer $g$, the $l_i'$ are nonnegative ordered integers at least
$g$ units apart, and (\ref{stunning}) gives the energies of a set of $n$ noninteracting particles on a standard one-body oscillator spectrum
but with their possible excitations $l_i^{\prime}$ separated by at least $g$ units, realizing $g$-exclusion\footnote{As a remark \cite{Ouvrybis,Dasbis} 
we note  that the very same $\beta\omega \to 0$ thermodynamic limit 
prescription used for the LLL-anyon thermodynamics (\ref{thermo}, \ref{bon}), specifically in the Calogero case ${1\over \sqrt{n}\beta\omega}\to {L\over \lambda}$ where $L$ is the length of the 1d line, would lead
to thermodynamics with a similar form
\[\ln Z_g = \int_0^{\infty}\rho_0(\epsilon)\ln y\; d\epsilon \;\;\;\oplus\;\;\;
y^g - y^{g-1} = z\rme^{-\beta\epsilon} \] 
where
$\rho_0(\epsilon)=L/(h  \sqrt{\epsilon/2m})$ is the  free  1d one-body density of state.
 This is yet another example, like the LLL-anyon model, of an $n$-body microscopic quantum Hamiltonian with thermodynamics of the type (\ref{newg}).
 Further, focusing on the $n$-body  LLL-anyon eigenstates  in the presence of a one-body harmonic well potential and their corresponding spectrum   (\ref{harmonic}), one can show \cite{Polybis} that,  in the vanishing magnetic field limit where (\ref{harmonic}) becomes the 1d harmonic well Calogero spectrum (\ref{top}),
 the LLL-anyon model can be explicitly mapped to the Calogero model. 
}.
From the $n$-body Calogero spectrum  (\ref{top}) the $n$-body partition function is obtained as
\[Z_{n}=x^{n/2+g n(n-1) / 2} \sum_{0 \leq l_i \leq l_{i+1}} x^{\sum_{i=1}^n l_i}=x^{n/2+g n (n - 1)/2} \prod_{j=1}^n{1 \over 1 - x^j}\]
which coincides with the $g$-exclusion partition function that one would obtain using (\ref{sala}) for the same one-body harmonic spectrum,
that is, $\epsilon_i = \omega(i+1/2)$, $i=0,1,2,\dots$ with $k \to \infty$.

To move to the ``Calogero-induced" $(1-g)$-inclusion, we apply the duality recipe: the $(1-g)$-grand partition function 
 is the inverse  of the Calogero $g$-exclusion grand partition function built from the above $Z_n$ with $z\to -z$.
We can in turn expand the resulting grand partition function in powers of $z$ 
to obtain the  $(1-g)$-inclusion  $n$-body  partition functions 
\[Z_{n}=x^{{n/2}+(1-g) n(n-1) / 2} \sum_{(g-1)(i-1) \leq l_i \leq l_{i+1}} x^{\sum_{i=1}^n l_i}\]
which coincide, as they should, with the $(1-g)$-inclusion partition functions 
one would obtain using (\ref{bibi}) for the {same} one-body harmonic spectrum.
These are, again, surprisingly simple expressions considering the somehow intricate construction leading to them.

However, there is even  more to contemplate:
the $(1-g)$-inclusion  $n$-body partition functions just obtained admit an interpretation in terms of the $n$-body Calogero-like  spectrum
\be{{\label{toto}}}
E_n=\omega\Big(\sum_{i=1}^n l_i+(1-g) \frac{n(n-1)}{2}+{n\over 2}\Big)\quad\quad (g-1)(i-1) \leq l_i \leq l_{i+1} \ee
which, we stress, is  not the naive $g\to 1-g$ mapping of the $n$-body Calogero spectrum (\ref{top}).
 In terms of the quasi-momentum $l'_i=l_i+(1-g)(i-1)$ it rewrites as
\be\label{stunningbis}E_n=\omega\big(\sum_{i=1}^n l_i^{\prime}+{n\over 2}\big) \quad\quad 0\leq l_i^{\prime} \leq l_{i+1}^{\prime}-(1-g)\ee
and for integer $g$ the $l_i'$ are integers obeying the ``inclusion" relation stated in (\ref{stunningbis}). Again, this is a stunningly simple result
for this ``Calogero-induced" inclusion $n$-body  spectrum.

The spectrum (\ref{toto},\ref{stunningbis}) begs for a Calogero-like Hamiltonian that would reproduce it. We have no such explicit Hamiltonian as yet,
but can make some remarks and offer speculation. The Calogero Hamiltonian (\ref{caloha}) is invariant under the duality
mapping $g \leftrightarrow 1-g$. Its spectrum (\ref{top}) and associated wavefunctions, however, are not. The reason is that wavefunctions must behave as
$x^g$ or $x^{1-g}$ near particle coincidence points, with $x$ their relative coordinate, and non-singularity and normalizability impose choosing the
positive power $g$, fixing the spectrum (\ref{top}). If the wavefunctions with behavior $x^{1-g}$ were somehow the relevant states of the Hamiltonian, then
the spectrum (\ref{toto}) would emerge, but still without the extra condition $(g-1)(i-1) \leq l_i$ which ensures its positivity.

One can speculate that the alternative set of wavefunctions with behavior $x^{1-g}$ are favored because of the existence of additional contact-type
interactions between the particles, which would both regularize their behavior within a small ``core" and make them the true states of the Hamiltonian.
The condition $(g-1)(i-1) \leq l_i$, on the other hand, ensures that the scaling of the states under an overall dilatation
$x_i  \to  \Lambda\, x_i$ for $\Lambda \to 0$, which is
\be
\psi \to \Lambda^{(1-g){n(n-1)\over 2} + \sum_i l_i}\, \psi
\nonumber\ee
has a positive exponent of $\Lambda$, indicating nonsingular behavior under a uniform shrinking of the coordinates. This should also arise as a dynamical condition.
The construction of a Hamiltonian with appropriate contact or other interactions
and the desired properties to reproduce the inclusion spectrum remains an interesting open problem.

\vspace{1cm}
\noindent{\bf Conclusions:} 

Inclusion statistics emerges as a natural extension of generalized quantum statistics in the domain where particles become more ``inclusive" 
or gregarious than bosons. The properties of systems with this type of statistics are quite intuitive, leading to relatively simple formulae for the energy
spectra in situations such as  LLL-anyons or Calogero particles, and are related by a remarkable duality relation to those of particles with exclusion statistics,
mapping  $g$-exclusion statistics to $(1-g)$-inclusion statistics.

A stunning property of free particles with inclusion statistics is that in low temperatures they condense  more readily than ordinary bosons \cite{bibibis},
for any statistics more inclusive than bosons. In particular, they condense in 2 dimensions, and marginally do not in 1 dimension, in contrast to bosons
which condense in 3 dimensions and marginally do not in 2 dimensions. The critical condensation temperature of inclusion particles is also higher than the one for
bosons, in situations where they both condense.

Preliminary numerical results on the average occupation number of states in a discrete spectrum reveal an even more intriguing pattern:
in ordinary Bose condensation, excited states have an average occupation as given by the grand canonical ensemble,
while the ground state is occupied by a macroscopically large number of particles, which form a superfluid. Inclusion particles, on the other hand,
condense through an enhanced occupation of {several} low-lying states sharing the particles in the condensate, creating a ``hump" in the occupation
numbers of low-lying states. The ground state plays a lesser role, maximum occupation arising for one of the excited states, and the condensate
appears as a superposition of several layers of superfluids. The full qualitative and quantitative properties of inclusion condensates are the subject of ongoing
investigation.

The Calogero model maintains its privileged position in this new statistical arena. Just as it can implement exclusion statistics in its regular manifestation,
it can also realize inclusion statistics through a simple and modest  alteration of its energy eigtenstates, which amounts to accepting some of the states rejected in the standard
Calogero model because of their singular behavior at particle coincidence points. As we speculated, this could be achieved
by introducing additional contact interactions between particles, effectively regularizing the appropriate set of singular wavefunctions. The
dynamical realization of this scenario through an explicit Hamiltonian remains a very interesting topic.

The enhanced propensity for condensation and the qualitatively different properties of the condensate compared to bosons make inclusion particles
interesting entities for theoretical study, but also for experimental work in situations where they can be realized as the effective statistics of
otherwise ordinary particles, such as the expanded ``inclusion" Calogero model. No doubt, inclusion statistics particles hold quite a few surprises and
interesting twists for those willing to explore them.

\vspace{1cm}
\noindent{\bf  Appendix:}

 {\bf Claim:} 
 one can alternatively obtain the exclusion and inclusion  $n$-body partition functions (\ref{sala}) and (\ref{bibi}) derived for the textbook one-body spectrum $\epsilon_1,\ldots,\epsilon_k$  directly from the   quantum numbers of  the Calogero and Calogero-induced spectra (\ref{stunning})  and (\ref{stunningbis}): indeed redefining the summation labels in (\ref{sala})  we see that
 it can be rewritten \`a la  (\ref{stunning}) as
 \[Z_n=\sum_{1+(i-1) g \leq l_i^{\prime} \leq {\text{min}}[l_{i+1}^{\prime}-g,k]} \exp(-\beta \sum_{i=1}^k\epsilon_{l_i^{\prime}})\]
Similarly  (\ref{bibi}) rewritten \`a la  (\ref{stunningbis}) becomes
 \[Z_n=\sum_{1 \leq l_i^{\prime} \leq {\text{min}}[l_{i+1}^{\prime}-(1-g),k]} \exp(-\beta \sum_{i=1}^k\epsilon_{l_i^{\prime}})\]
 
 {\bf Remark:} in the canonical ensemble where the number of particles $n$ is fixed the $(1-g)$-inclusion  multiplicities (\ref{mult})  induce  a new probability distribution on the $n_i$'s: for a configuration $\{n_1,n_2,\ldots,n_k\}$  such that 
 $\sum_{i=1}^k n_i=n$
 \[P(n_1,n_2,\ldots,n_k)={{{{\text {m}}_{1-g}(n_1,\ldots,n_k)}(\rme^{-\beta\epsilon_1})^{n_1}\ldots(\rme^{-\beta\epsilon_k})^{n_k}}\over \sum_{n_1, \ldots, n_k=0,\;\sum_{i=1}^k n_i=n
}^{\infty}{{\text {m}}_{1-g}(n_1,\ldots,n_k)}(\rme^{-\beta\epsilon_1})^{n_1}\ldots(\rme^{-\beta\epsilon_k})^{n_k}}\]
in place of the standard canonical  Bose distribution i.e.,
 \[P(n_1,n_2,\ldots,n_k)={{(\rme^{-\beta\epsilon_1})^{n_1}\ldots(\rme^{-\beta\epsilon_k})^{n_k}}\over \sum_{n_1, \ldots, n_k=0,\;\sum_{i=1}^k n_i=n
}^{\infty}(\rme^{-\beta\epsilon_1})^{n_1}\ldots(\rme^{-\beta\epsilon_k})^{n_k}}\]

{{\bf Remark:} 
We note that the  Calogero-induced  $(1-g)$-inclusion $n$-body spectrum (\ref{toto})  happens to generate the permutations  of  integers  in $n$ parts $l'_1,l'_2,\ldots, l'_n$ such that for example  the $i$-th part $l'_i$  is strictly greater that $(g-1)i$ (these integers are then necessarily   larger that $(g-1) {n(n+1)}/{2}+n $) . This can be directly seen by considering instead of (\ref{toto}) the equivalent spectrum  up to a global shift  
\be\nonumber E'_n=\omega\big(\sum_{i=1}^n l_i+ n\; g\big)\quad\quad (g-1)(i-1) \leq l_i \leq l_{i+1} \ee 
in turn equivalent to 
\be\nonumber E'_n=\omega \sum_{i=1}^n l'_i \quad\quad (g-1)i < l'_i \leq l'_{i+1} \ee 
 whose degeneracy gives the number of such said permutations (for $g=2\to (1-g)=-1$ see in particular the Online Encyclopedia Integer Sequences A244239-46).

\end{document}